\documentclass[aps,prx,floatfix,twocolumn,notitlepage,superscriptaddress,10pt]{revtex4-2}

\usepackage{xcolor}
\usepackage{grffile}
\usepackage{amsmath, amsthm, amssymb, bbold}
\usepackage[normalem]{ulem}
\usepackage{cancel}
\usepackage{bm}
\usepackage{microtype}
\usepackage{hyperref}
\usepackage{setspace}
\usepackage{grffile}
\hypersetup{colorlinks,linkcolor=blue,urlcolor=blue,citecolor=blue}
\usepackage{amsmath}    
\usepackage{amssymb}
\usepackage{graphicx}   
\usepackage{verbatim}   
\usepackage{color}      
\usepackage{hyperref}   
\usepackage[normalem]{ulem}
\usepackage{natbib}
\usepackage{enumitem} 
\usepackage{dsfont} 
\usepackage{physics} 
\usepackage{lipsum}	

		\newcommand{\be}{\begin{equation}}
		\newcommand{\ee}{\end{equation}}

\begin{document}

\title{Independent stabilizer R\'enyi entropy and entanglement fluctuations in random unitary circuits}

    \author{Dominik Szombathy}
	   \email{szombathy.dominik@edu.bme.hu}
	   \affiliation{Department of Theoretical Physics,  Institute of Physics, Budapest University of Technology and Economics,  M\H{u}egyetem rkp. 3., H-1111 Budapest, Hungary }
	   \affiliation{Nokia Bell Labs, Nokia Solutions and Networks Kft, 1083 Budapest, B\'okay J\'anos u. 36-42, Hungary}
    \author{Angelo Valli}
        \affiliation{Department of Theoretical Physics,  Institute of Physics, Budapest University of Technology and Economics,  M\H{u}egyetem rkp. 3., H-1111 Budapest, Hungary }
        \affiliation{MTA-BME Quantum Dynamics and Correlations Research Group, Budapest University of Technology and Economics,  M\H{u}egyetem rkp. 3., H-1111 Budapest, Hungary}
   \author{C\u at\u alin Pa\c scu Moca}
      \affiliation{MTA-BME Quantum Dynamics and Correlations Research Group, Budapest University of Technology and Economics,  M\H{u}egyetem rkp. 3., H-1111 Budapest, Hungary}
        \affiliation{Department of Physics, University of Oradea,  410087, Oradea, Romania}
    \author{L\'or\'ant Farkas}
        \affiliation{Nokia Bell Labs, Nokia Solutions and Networks Kft, 1083 Budapest, B\'okay J\'anos u. 36-42, Hungary}
    \author{Gergely Zar\'and}
          \email{zarand.gergely.attila@ttk.bme.hu}
        \affiliation{Department of Theoretical Physics,  Institute of Physics, Budapest University of Technology and Economics,  M\H{u}egyetem rkp. 3., H-1111 Budapest, Hungary }
       \affiliation{MTA-BME Quantum Dynamics and Correlations Research Group, Budapest University of Technology and Economics,  M\H{u}egyetem rkp. 3., H-1111 Budapest, Hungary}
\date{\today}

\begin{abstract}
We investigate numerically the joint distribution of magic ($M$) and entanglement ($S$) in 
$N$-qubit Haar-random quantum states. 
The distribution $P_N(M,S)$ as well as the marginals become exponentially localized, and centered around the values 
$\tilde{M_2} \to N-2$ and $\tilde{S} \to N/2$ as $N\to\infty$. 
Magic and entanglement fluctuations are, however, found to become exponentially uncorrelated. 
Although exponentially many states with magic $M_2=0$ and entropy $S\approx S_\text{Haar}$ exist, 
they represent an exponentially small fraction compared to typical quantum states, which are 
characterized by large magic and entanglement entropy, and uncorrelated magic and entanglement fluctuations.
\end{abstract}

\maketitle

\section{Introduction}

Although it may still be considered somewhat elusive, quantum complexity is a central theme 
in quantum computing.
Certain 'simple' quantum states are in some sense close to classical, while others are 'complicated' 
as they are strongly non-local and possess quantum correlations between remote components or far-away states of a system.  
To distinguish 'simple' and 'complicated'  states,  several tools and concepts such as entanglement, non-locality, quantum 
coherence, or quantum correlations, have been proposed 
to quantify quantum complexity in the framework of quantum resource theory~\cite{Brandao2013,veitch2014resource,Chitambar2019,Bauer2020}. 

In this context, although difficult to measure~\cite{Guhne2007,Guhne2009,Horodecki2009}, \emph{entanglement}  
is of utmost significance, and has long been identified as one of the major quantum resources. 
Entanglement entropy helps to characterize the amount of classical resources needed to 
represent a state, and entanglement is the key ingredient of quantum teleportation~\cite{Furusawa1998,Pirandola2015,bouwmeester1997experimental,Hu2023}, 
quantum repeaters~\cite{Briegel1998,Munro2015,Azuma2023}, as well as many quantum algorithms~\cite{cleve1998quantum,Montanaro2016}. 

Entanglement and non-locality, however, do not reflect the entire complexity of a state. 
Another way to characterize the complexity of a  state $|\psi\rangle$ 
is to investigate the complexity of the quantum circuit that produces this state.
While generic quantum dynamics typically turns simple product states 
 into states that are impossible to represent on a classical computer~\cite{jaeger2007classical}, 
yet highly entangled states can be non-complex in this sense. 
A striking example is given by entangled quantum states produced 
by classically efficiently simulable Clifford circuits~\cite{Gottesman1998, aaronson2004improved,nahum2017quantum,kim2013ballistic,chandran2015semiclassical,znidaric2020entanglement}. 

The lack of complexity of these states and the corresponding circuits can be captured    
through the \emph{spectral property} of the Clifford-circuits~\cite{szombathy2024}, or by 
the notion of  \emph{magic} or \emph{non-stabilizing power}. 
The concept of non-stabilizing power, in particular, is based on the notion 
of  \emph{stabilizer Rényi entropy} (SRE)~\cite{Leone2022}. 
The latter quantifies the hardness to  simulate classically a quantum state in terms of 
simple stabilizer states~\cite{Huang2024},  and has raised significant attention 
recently~\cite{Oliviero2022,Haug2023,Rattacaso2023,Haug2024,Leone2024,Turkeshi2024,Turkeshi2407.03929,Ahmadi2024,Haug2024, Fux2024,tarabunga2024critical, zhang2024quantum,mello2024retrieving,turkeshi2023measuring,Lami2023,Tarabunga2024,tarabunga2024magic,piemontese2023entanglement}.

In a recent work, a distinction between entanglement-dominated and 
magic-dominated states was identified, suggesting a division of the Hilbert space into two distinct regimes~\cite{gu2024}. The interplay between magic and entanglement has been further explored 
in the context of matrix product states, particularly in a spin-1 model, by examining 
the relationship between magic and the bond dimension required to approximate the ground 
state~\cite{Frau2024}. An even more ambitious study investigated the comparison between 
many-body entanglement and magic in nuclear structure~\cite{brokemeier2024}.
The relationship between these two quantities remains ambiguous: it is possible to find states with negligible magic but high entanglement, as well as states with low entanglement yet substantial magic. Ground states, in this context, stand out as atypical, as they tend to exhibit reduced entropy compared to more typical states.
However, entanglement and magic are related, as entanglement is \textit{instrumental} to generate high-magic states. 
Indeed, the $N$-qubit state with the highest magic that can be generated \textit{without} entanglement 
is given by $\ket{\psi} = \bigotimes_{j=1}^{N} T_j \ket{0}_j$, 
where $T$ is the single-qubit $\pi/4$ phase gate. 
The corresponding magic, measured by the SRE is therefore $M_2(\ket{\psi}) =  N \times 0.585...$~\cite{szombathy2024}, 
which is subtantially lower than the theoretical upper bound of the SRE, $\log_2((2^N+1)/2)$~\cite{Leone2022}. 

In this work, we address the question, if magic and entanglement are related in generic, Haar-random quantum states. 
First, we study numerically the distribution of magic, $P_N(M_2)$ of few-qubit quantum systems. 
We find that for Haar-random states of an $N$-qubit circuit, $P_N(M_2)$ is an exponentially sharp distribution 
centered at around $\tilde{M}_2 \approx N-2$. 
This follows also from the concentration property of the linear magic $M_{lin}$, 
which is a consequence of Levy's lemma~\cite{Leone2022}. 
The entanglement entropy of Haar-random $N$-qubit states displays a similar, exponentially sharp distribution
around $\tilde{S}$,  approaching $N/2$ in the large $N$ limit. 
The typical values of entropy and magic are therefore strongly correlated.
However, their joint distribution, $P_N(M_2,S)$ becomes uncorrelated 
for large $N$'s, 
and the covariance of the entropy and magic fluctuations vanish exponentially fast. 
 
The manuscript is structured as follows: In Sec.~\ref{sec:SRE}, we examine the stabilizer R\'enyi entropy (magic) 
and present results for its distribution for various system sizes. 
In Sec.~\ref{sec:S}, we focus on entanglement, 
while Sec.~\ref{sec:Joint} explores the relationship between magic and entanglement, 
along with results for their joint distribution. 
Finally, we conclude in Sec.~\ref{sec:C} with a summary of the findings.

\section{Stabilizer R\'enyi Entropy (SRE) distribution of Haar-random states} 
\label{sec:SRE}

Stabilizer R\'enyi entropy (SRE), also referred to as magic, represents the hardness to classically simulate a 
quantum state~\cite{Leone2022,haug2023stabilizer,Leone2024}  within the stabilizer formalism. This entropy is closely related to 
the structure of Clifford circuits, i.e. unitary operations mapping Pauli $N$-strings, $\sigma\in {\cal P}_N \equiv  \{1, X, Y,Z\}^N $,  
to  $N$-strings – apart from an overall sign. 
Each state $|\psi\rangle\langle \psi|$ can be decomposed in terms of  Pauli strings, 
and one can associate with it a corresponding string probability  distribution
\begin{equation}
\Xi_\psi(\sigma) = \frac 1 d  \bra{\psi} \sigma \ket{\psi}^{2},
\end{equation}
where $d=2^N$ is the dimension of the Hilbert space of $N$ qubits. Here we focus on the 
corresponding stabilizer 2-R\'enyi entropy, which we henceforth refer to as SRE or magic, 
\begin{equation} 
   M_2 (\ket{\psi}) \equiv  - \log_2 \sum_{\sigma \in \mathcal{P}_N} \Xi^2_\psi(\sigma) - \log_2 (d), 
   \label{eq:SRE}
\end{equation}
with $\mathcal{P}_N$ the set of unsigned Pauli strings. 
SRE fulfills the following important  properties~\cite{Leone2022}:
(i) $M_2$ remains unaltered  under Clifford operations; 
(ii) $M_2(\ket{\psi}) = 0$  iff  $\ket{\psi}$ is a stabilizer state;
(iii) similar to entropy, $M_2$ is additive for product states; 
(iv) it is upper bounded  by $M_2(\ket{\psi}) \leq \log_2((2^N+1)/2)$. 

 \begin{figure}[t!]
    \includegraphics[width=\linewidth]{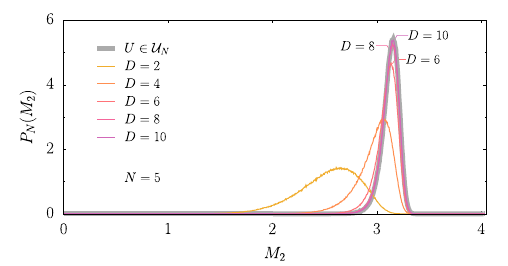}
    \caption{Convergence with circuit depth. 
    Distribution of magic $P_N(M_2)$ generated by random unitary operators $U \in \mathcal{U}_N$ for $N=5$. 
    In a brickwall construction with only 2-qubit gates, the distribution converges 
    to the one generated by a single unitary operator at circuit depth $D \approx 2N$. }
    \label{fig:M2_brickwall}
\end{figure}

\begin{figure}[t!]
    \includegraphics[width=\linewidth]{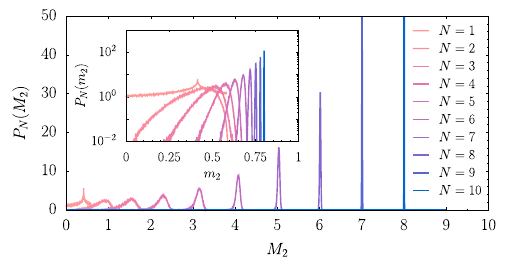}
    \caption{Distribution of magic $P_N(M_2)$ generated by random unitary operators $U \in \mathcal{U}_N$. 
    Increasing $N$, the distribution becomes concentrated around the value $\tilde{M}_2 \approx N-2$. 
    Inset: distribution of magic density $m_2=M_2/N$. }
    \label{fig:M2_distribution}
\end{figure}

In the following, we show the distribution of magic $P_N(M_2)$, obtained by sampling random circuits $U$ acting 
on the reference state $\ket{0} \equiv \bigotimes_{j=1}^N \ket{0}_j$ of an $N$-qubit register, 
i.e., $\ket{\psi} = U \ket{0}$. 
Specifically, Fig.~\ref{fig:M2_brickwall} considers the case of a brick-wall circuit $U_{\mathrm{BW}}$ of depth $D$, 
constructed by alternating even and odd layers of 2-qubit Haar-random unitaries. 
The corresponding magic distribution converges rapidly to that of Haar-random circuits $U \in \mathcal{U}_N$,  
and for a circuit depth $D\gtrsim 2N$, the distributions obtained sampling $U_{\mathrm{BW}}$ and $U$ are nearly indistinguishable. 
For the remained of this work we focus on the distribution of generic Haar-random operators $U \in \mathcal{U}_N$, and 
Fig.~\ref{fig:M2_distribution} displays the evolution of $P_N(M_2)$ with increasing number of qubits. 
The case $N=1$ is somewhat peculiar~\cite{szombathy2024}; the saddle point states  
on the Bloch sphere give rise to logarithmic van Hove singularities. 
For larger values of $N$, however, the distribution gradually shifts to higher values, 
and is concentrated 
at around $\tilde{M}_2 \approx N-2$. 
The numerically observed maximum is extremely close to the lower bound, 
$\mathbb{E} [{M}_2(\psi_\text{Haar})] \geq N - 2 + \log_2(1+3/2^N)$~\cite{Leone2022}.  

The logarithmic inset of Fig.~\ref{fig:M2_distribution} shows the magic density, and 
demonstrates that the distribution becomes exponentially narrow 
with increasing $N$, 
and the magic density of typical Haar-random states converges to $m_2 \approx 1-2/N \to 1$ in the large $N$ limit, 
as dictated by the bounds reported earlier. 

\begin{figure}[t!]
    \includegraphics[width=\linewidth]{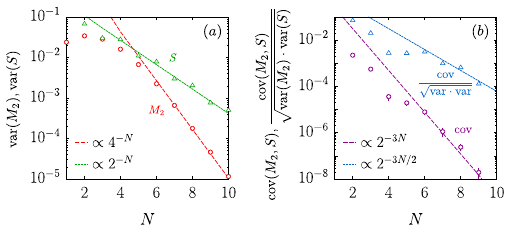}
    \caption{(a) Scaling of the variance of the magic (red circles) and entanglement (green triangles) distributions. 
    Both are exponentially vanishing in the limit $N \to \infty$. 
    (b) Scaling of the covariance of the joint distribution $P_N(M_2,S)$ (magenta circles) 
    which is exponentially suppressed with system size, also faster that the geometric mean of the individual variances (blue triangles), 
    demonstrating that magic and entanglement fluctuations are asymptotically uncorrelated. 
    The color dashed lines highlight the corresponding exponential scaling for each quantity. }
    \label{fig:variance}
\end{figure}

The width of the distribution scales as $\delta M_2 =\sqrt{\textrm{var}(M_2)} \approx 2^{-N}$ (see also Fig.~\ref{fig:variance}). 
This behavior is in contrast to standard thermodynamic behavior, which would 
correspond to a variance scaling as $\sim N$.
The ultimate reason behind this behavior is that, apart from an overall phase and the normalization, 
a Haar-random states are characterized by 
exponentially many ($\sim 2^N$) complex numbers, and the sum in Eq.~\eqref{eq:SRE}
can therefore be considered as a sum of $N_\sigma = 4^N$ weakly correlated terms. 
A simple calculation yields $\delta M_2 \sim 1/N_\sigma^{1/2}\sim 2^{-N}$, as we indeed observe it numerically. 
A similar result holds for the linear magic~\cite{Leone2022}. 

Although the magic distribution $P_N(M_2)$ is strongly skewed for small $N$'s, 
we find that higher-order cumulants vanish faster than the variance, 
i.e., $\kappa_{n>2}/\kappa_2^{n/2} \to 0$, 
and in the thermodynamic limit the distribution converges towards a Gaussian with a vanishing variance
$P_{N \to \infty}(M_2) \to \mathcal{N}(\kappa_1 \to \tilde{M}_2, \kappa_2 \to 0)$.

\begin{figure}[t!]
    \includegraphics[width=\linewidth]{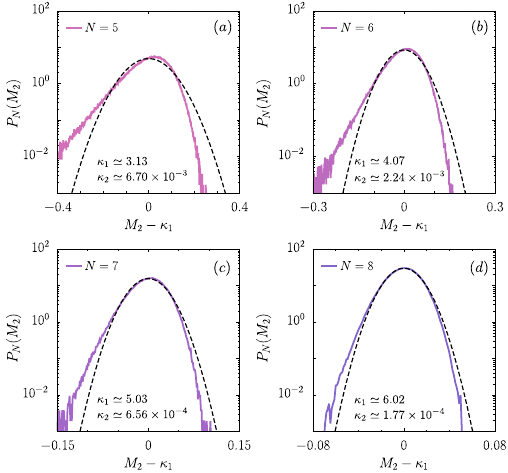}
    \caption{Distribution of magic $P_N(M_2)$ centered around the first cumulant $\kappa_1=\langle M_2 \rangle$ on a semilog scale. 
    Black dashed lines correspond to a normal distribution 
    with mean $\kappa_1$ and variance $\kappa_2=\langle (M_2 - \langle M_2\rangle)^2 \rangle$ extracted from $P_N(M_2)$.  }
    \label{fig:M2_logrho}
\end{figure}

\begin{figure}[t!]
    \includegraphics[width=\linewidth]{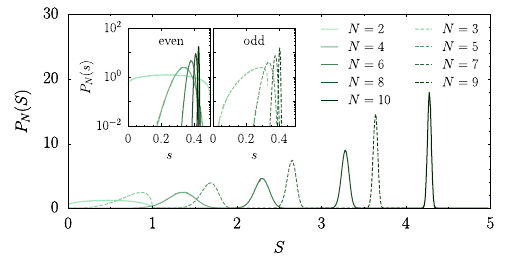}
    \caption{Distribution of entanglement entropy $P_N(S)$ generated by random unitary operators $U \in \mathcal{U}_N$. 
    Increasing $N$, the distribution becomes concentrated around the value $\tilde{S}_2 \approx N/2$.
    Insets: distribution of entropy density $s=S/N$. }
    \label{fig:S_distribution}
\end{figure}

\begin{figure}[t!]
    \includegraphics[width=\linewidth]{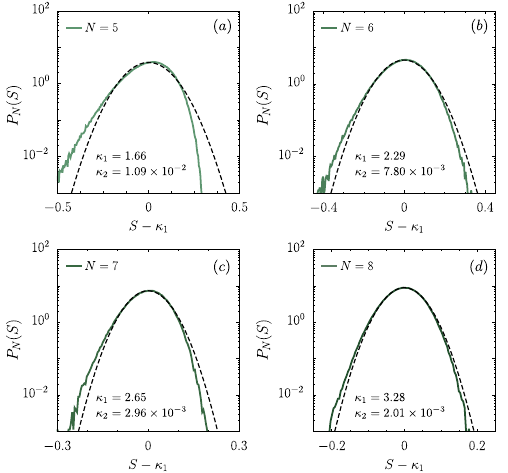}
    \caption{Distribution of magic $P_N(S)$ centered around the first cumulant $\kappa_1=\langle S \rangle$ on a semilog scale. 
    Black dashed lines correspond to a normal distribution 
    with mean $\kappa_1$ and variance $\kappa_2=\langle (S - \langle S\rangle)^2 \rangle$ extracted from $P_N(S)$.  }
    \label{fig:S_logrho}
\end{figure}

\section{Entanglement entropy distribution} 
\label{sec:S}

Entanglement entropy displays a behavior very similar behavior to that of magic. 
We have computed the von Neumann entropy of Haar-random states by 
cutting the system to two parts of equal size, $N/2$ for even $N$,  
and of almost equal size, $(N \pm 1)/2$ for odd $N$.
The evolution of entropy distributions $P_N(S)$ is presented in Fig.~\ref{fig:S_distribution}.  
Also in for the entanglement entropy, we observe an exponentially narrow 
distribution, 
concentrated around $\tilde{S} \approx N/2 - \log_2(e)/2$ for even $N$, 
and around $\tilde{S} \approx (N + 1)/2 - \log_2(e)$ or $\tilde{S} \approx (N - 1)/2 - \log_2(e)/4$ 
for odd $N$, for the subsystem with $(N \pm 1)/2$ qubits, 
consistent with the Page formula~\cite{page1993average,odavic2412.10228}. 
In contrast, the probability of maximally entangled states is very small~\cite{Nadal2011}. 
The distribution of the rescaled entropy, $s = S/N \approx 1/2 - \log_2(e)/(2N)$, 
shown in the inset of Fig.~\ref{fig:S_distribution}, gradually shifts toward $\tilde{s} = 1/2$, 
which corresponds to the entropy density of a state at infinite temperature.

\begin{figure*}[ht!]
    \includegraphics[width=\textwidth]{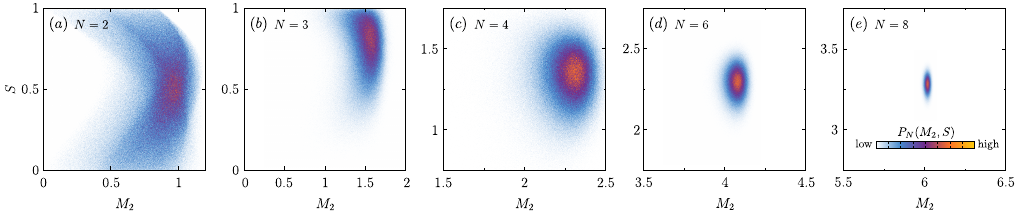}
    \caption{
    Joint distribution $P_N(M_2,S)$ of magic and entanglement entropy. 
    Since the variance of magic and entropy vanish at a different rate, the joint distribution assumes a characteristic ellipsoidal shape 
    as it concentrates around $\tilde{M}_2$ and $\tilde{S}$. }   
    \label{fig:joint_distribution}
\end{figure*}

The observed trend can be explained in terms of the eigenstate thermalization hypothesis~\cite{Deutsch2018}. 
Generic quantum states of a large system produce a thermal state when projected to a subsystem. 
Infinite temperature states of a subsystem have the largest entropy due to the fact that 
they are exponentially abundant when sampling states randomly in the Hilbert space~\cite{hamma2012quantum,zhou2019emergent}. 
For generic quantum states, the entropy of a subsystem approaches $N/2 - \log_2(e)/2$ with increasing $N$, 
though finite-size corrections introduce deviations.


Similarly to magic, entanglement distributions become exponentially narrow with increasing $N$.
This can be understood similarly to the sharpness of magic distributions.  
Entanglement is an average over all $2^M$ density matrix eigenvalues  
of an $M$-qubit subsystem. Assuming weak correlations between these, one therefore 
arrives at entropy fluctuations of size $\delta S =\sqrt{\textrm{var}(S)}\approx 2^{-N/2}$, 
as indeed verified by the quantitative analysis presented in the 
next section and in agreement with previous results~\cite{zhou2019emergent}.

\section{Joint distribution and correlations} 
\label{sec:Joint}

Let us now turn to the main subject of this work, the analysis of joint magic and entanglement 
entropy distributions of Haar-random states. Fig.~\ref{fig:joint_distribution} displays the 
evolution of the joint distribution with increasing qubit number. 

In Fig.~\ref{fig:joint_distribution} we display the joint distribution $P_N(M_2,S)$. 
Clearly, for small values of $N=\{1,2\}$, $P_N(M_2,S)$ displays a particular structure as is visible in Fig.~\ref{fig:joint_distribution}(a,b).
As the system size $N$ increases, the averages of the rescaled magic and entanglement remain fixed at 
$\tilde{M}_2 \approx N-2$ and $\tilde{S} \approx N/2$, respectively, while their variances exhibit rapid exponential decay: 
$\textrm{var}(M_2) \propto 4^{-N}$ and $\textrm{var}(S) \propto 2^{-N}$.  
These findings are in excellent agreement with the simple estimates derived in previous sections, 
as corroborated by the numerical data presented in Fig.~\ref{fig:variance}(a). 
Notably, the numerical results are best described by exponential decay, confirming the robustness of the predictions.  

In Fig.~\ref{fig:variance}(b), we also present the covariance between the two quantities, $M_2$ and $S$, 
which follows an exponential decay with a slightly larger exponent: $ \textrm{cov}(M_2, S) \propto 2^{-3N} $. 
Importantly, the covariance decays faster that the geometric mean of the variance of $M_2$ and $S$. 
This 
indicates that the correlation between magic and entanglement becomes negligible as $N \to \infty$. 
This behavior strongly supports the conclusion that, in the thermodynamic limit, the two quantities become effectively uncorrelated. 
Recent results from Hamma and collaborators~\cite{Hamma_unpublished} 
show analytically that for the linear magic and entanglement entropy, the correlation vanish \textit{identically} for any $N$. 
Our numerical results extends and generalize this for $M_2$ and the von Neumann entropy, 
showing that the correlation vanish exponentially with system size. 

\section{Conclusions}\label{sec:C}

In this work, we investigated the joint distribution of stabilizer Rényi entropy 
(magic) and entanglement entropy in Haar-random quantum states of $N$ qubits, 
uncovering key statistical properties of these quantum resources. 
Our numerical analysis revealed that both magic and entanglement exhibit exponentially
sharp distributions as the system size increases, with magic centering around $N-2$ 
 and entanglement entropy peaking near $N/2$. While these quantities are strongly 
 correlated in their typical values, their fluctuations become exponentially uncorrelated 
 in the thermodynamic limit, as demonstrated by the rapid decay of their covariance. 
 For the linear magic, it can be shown analytically that the covariance identically vanishes~\cite{Hamma_unpublished}. 
 This result emphasizes the fundamental independence of these two measures of quantum complexity in 
 large systems, even though they individually quantify distinct aspects of quantumness.

Additionally, our work demonstrates that  quantum states with vanishing magic but
 high entanglement entropy, or states with high magic but low entanglement, are 
 exponentially rare compared to the abundance of states characterized by both high 
 magic and high entanglement, indicating the predominance 
 of highly complex quantum states in the Hilbert space, as 
 most generic states are characterized by large values of both resources
 
 Our findings highlight the utility of random unitary circuits as a framework for exploring quantum resource theories, particularly for understanding the interplay and statistical properties of key measures such as magic and entanglement.

\acknowledgments
This research was supported by the Ministry of Culture and Innovation and the National Research, Development and Innovation Office (NKFIH) within the Quantum Information National Laboratory of Hungary
(Grant No. 2022-2.1.1-NL-2022-00004), through NKFIH research grants No. SNN139581, 
and QuantERA `QuSiED' grant No. 101017733.
D.S. acknowledges the professional support of the doctoral student scholarship program of the co-operative doctoral program of the Ministry for Innovation and Technology from the source of the National Research, Development and Innovation fund.
C.P.M. acknowledges support by the Ministry of Research, Innovation and Digitization, CNCS/CCCDI–UEFISCDI, under the projects PN-IV-P1-PCE-2023-0159  and PN-IV-P1PCE-2023-0987.

\bibliography{references}

\end{document}